\documentclass{aa}

\usepackage{txfonts}
\usepackage[ansinew]{inputenc}
\usepackage{textcomp}
\usepackage{graphicx}
\usepackage{latexsym}

\begin{document}

\title{Near-infrared polarimetry setting constraints on the orbiting spot model for Sgr~A* flares}

\author{L. Meyer\inst{1} \and A. Eckart\inst{1} \and R. Sch\"odel\inst{1} \and W. J. Duschl\inst{2,3} \and K. Mu\v{z}i\'{c}\inst{1}  \and M. Dov\v{c}iak\inst{4} \and V.~Karas\inst{4}}

\institute{I.Physikalisches Institut, Universit\"at zu K\"oln, Z\"ulpicher Str. 77, 50937 K\"oln, Germany \and Institut f\"ur Theoretische Physik und Astrophysik, Universit\"at Kiel, 24098 Kiel, Germany \and Steward Observatory, The University of Arizona, 933 N. Cherry Ave. Tucson, AZ 85721, USA \and Astronomical Institute, Academy of Sciences, Bo\v{c}n\'{i} II, CZ-14131 Prague, Czech Republic}

%
\abstract
{Recent near-infrared polarization measurements of Sgr~A*
 show that its emission is significantly polarized during flares and consists of a non- or weakly polarized main flare with highly polarized sub-flares. The flare activity suggests a quasi-periodicity of $\sim $20 minutes in agreement with previous observations.}
{By simultaneous fitting of the lightcurve fluctuations and the
time-variable polarization angle, we address the question of whether
these changes are consistent with a simple hot spot/ring model, in which the interplay of relativistic effects
plays the major role, or whether some more complex dependency of the
intrinsic emissivity is required.}  
{We discuss the significance of the 20\,min peak in the periodogram of a flare from 2003. We consider all general relativistic effects that imprint on the polarization degree and angle and fit the recent polarimetric data, assuming that the synchrotron mechanism is responsible for the intrinsic polarization and considering two different magnetic field configurations.} 
{ Within the quality of the
available data, we think that the model of a single spot in addition to an underlying ring is favoured. In this model the broad near-infrared flares of Sgr~A* are due to a sound wave that travels around the MBH once while the sub-flares, superimposed on the broad flare, are due to transiently heated and accelerated electrons which can be modeled as a plasma blob. Within this model it turns out that a strong statement about the spin parameter is difficult to achieve, while the inclination can be constrained to values $\ga 35\degr$ on a $3\sigma$ level. }
{} 


\titlerunning{Constraints on the orbiting spot model for NIR Sgr~A* flares}

\maketitle

\section{Introduction}

It is by now widely accepted that the compact radio source Sagittarius A* (Sgr~A*) is the manifestation of a massive (M $=3.6\cdot 10^6\textnormal{M}_{\sun}$) black hole (MBH) at the center of the Milky Way (e.g. Sch\"odel et al. \cite{rainer1}, \cite{rainer2}; Genzel et al. \cite{genzel}; Ghez et al. \cite{ghez}; Eckart et al. \cite{ecki1},b). It is the closest MBH, with a distance of only $\sim 7.6$ kpc (Eisenhauer et al. \cite{eisenhauer}). The first observation of its highly variable near-infrared (NIR) counterpart has been reported by Genzel et al. (\cite{genzel}). The short rise and fall timescales of these NIR flares point to physical processes within a few Schwarzschild radii ($R_S$) of the MBH. Simultaneous NIR/X-Ray observations showed that the high variability at these wavelengths have a common physical origin, with a spectrum pointing at synchrotron and synchrotron self-Compton models with rapid cooling of transiently heated electrons (e.g. Eckart et al. \cite{ecki3}, \cite{ecki1}; { Liu et al.~\cite{liu2}; Liu \& Melia~\cite{liu}}; Yuan et al. \cite{yuan1}, \cite{yuan2}; { Markoff et al. \cite{markoff}}).

In a few NIR and X-Ray flares a suggestive quasi-periodicity of roughly $\sim20$ minutes has been reported (Genzel et al. \cite{genzel}; Eckart et al. \cite{ecki2}; Belanger et al. \cite{belanger}). It manifests itself in the lightcurve as sub-flares superimposed on the underlying main flare (see Figure \ref{Fig:flare2003}). 
It is not clear yet, how these quasi-periodicities (QPOs) are created. The timescale of the lightcurve variations and the rather small volume around Sgr~A*, where these variations originate, indicate that rapid motion in strong gravity is involved. Frequencies in microquasar QPOs scale with the inverse mass of the BH, and the frequency of the QPOs in Sgr~A* indicates that an extrapolation from the stellar-mass BH case in microquasars to the MBH case is applicable (Abramowicz et al. \cite{abramowicz}; McClintock \& Remillard \cite{mcclintock} and references therein). 

The timescale of $\sim 20$ minutes is comparable to the orbital timescale $T$ near the innermost stable circular orbit (ISCO) of a spinning MBH (Bardeen et al. \cite{bardeen}). This can be written as 
\begin{equation}
\label{period}
T\doteq 110\, \left( r^{3/2}+a_{\star}\right)\, \frac{M}{3.6\cdot 10^6M_{\sun}}\; \textnormal{[sec]},
\end{equation}
where $a_{\star}$ is the BH dimensionless spin parameter ($-1\leq a_\star \leq 1$), and $r$ is a circumferential radius within the equatorial plane given in units of the gravitational radius $r_g\equiv GM/c^2$.
Therefore it appears reasonable that corresponding frequencies
are present in the flare lightcurves. In this context, we would also
like to point out the rapid variability in the light curve of the NIR flare of 16 June 2003 shown in Figure 1 (originally reported by Genzel et al. \cite{genzel}). Variations greater than $5\sigma$ can be seen on timescales of a few minutes, inferring an upper limit of the size of the source of less than 10~Schwarzschild radii if assuming that the cause responsible for the flare propagates at the speed of light. It can be expected, however, that the actual signal in the source propagates with a speed comparable to
the sound speed or to the Alfv\'en speed, i.e.\ a few orders of
magnitude slower than $c$. Therefore it appears again reasonable to assume
orbital motion of a compact source as the cause of the rapid variability. 

Furthermore, global MHD simulations show that the inner accretion flow is very inhomogenous and nonuniform, so that local overdensities build up (Machida \& Matsumoto \cite{machida}; DeVilliers et al. \cite{devilliers}). Modeling these overdensities as hot spots orbiting around the BH give lightcurves very similar to the observed ones (Broderick \& Loeb \cite{broderick1}, \cite{broderick2}; Eckart et al. \cite{ecki2}).  
Recent polarimetric measurements showed that the sub-flares have a degree of linear polarization of up to $\sim 20$\% and that they come along with a swing in the polarization angle (Eckart et al. \cite{ecki2}). The new information provided by the polarization state is extremely useful because it breaks the degeneracy of various model parameters. In this paper we demonstrate that the available data are consistent with an orbiting spot model and that they offer a new avenue to constrain BH parameters in the future.

In the next section, we first re-analyse the NIR flare observed by Genzel et al. (\cite{genzel}). Using the method of Vaughan~(\cite{vaughan}) we examine the significance of the peak in the periodogram at $\sim$20 min. After that, we address the question what the nature of the broad underlying main flare might be. We then use this, together with the hot spot model for the sub-flares, to fit the recent NIR polarimetric observations by Eckart et al.~(\cite{ecki2}). Details of the model and the fitting procedure are given. In Sect.~\ref{concl} we draw conclusions and discuss caveats, with special emphasis on the effects and analytic properties of higher order images.  

\section{Quasi-periodicity in NIR flares from Sgr A*?}

Up to now, quasi-periodicities in NIR flares from Sgr~A* have been
reported just for the three K-band flares described in
Genzel et al. (\cite{genzel}) and Eckart et al.~(\cite{ecki2}), {observed at the VLT. Their detection at longer wavelengths has not yet been reported on.
The reason for this may be indicative for a lower sub-flare
contrast at longer wavelengths due to intrinsic spectral properties of the source
and to the fact that at these wavelengths flux contributions from an extended
dust component complicates the determination of the flux density from Sgr~A*
(Eckart et al. \cite{ecki1}, Ghez et al. \cite{ghez2}). }Concerning the X-Ray regime, there seems to be also one fairly firm case of QPOs in a flare observed by XMM (Belanger et al. \cite{belanger}).

A serious limitation for the possible detection of quasi-periodicities in NIR
flares are observational constraints. Several factors must coincide: (i) the occurrence of a
flare; (ii) the flare must be fairly bright and last long enough in order
to be able to sample a sufficient number of oscillations; (iii) atmospheric
conditions must be good (seeing $\la 0.8\arcsec$ and $\tau_0 \ga 3$\,ms at NACO/VLT) and --
above all -- stable during the entire observation, i.e. several hours. We estimate from our observational experience that these conditions are fulfilled only 10-20\% of the total time dedicated to observations with NACO at the VLT. An additional
problem is caused by gaps in the observational sequence that may be introduced into the time series due to
the necessity of sky background measurements in NIR observations.  Due
to the extremely crowded field of the stellar cluster in the GC, the background
cannot be extracted from the on-source observations.  Also it is not clear whether each flare would be
accompanied by a QPO. A new window into QPOs may have been opened by the
discovery of Eckart et al. (\cite{ecki2}), who
found that possible periodic variations can be observed in polarized
light while not being detectable in the overall flux.

In the context of QPOs, it is also important to refer to the debate
concerning the existence of QPOs in AGN. In many cases X-ray light curves of these
objects appear to be characterised by red noise, which may lead to an
overestimation of the significance of periodic signals in the light
curves (see e.g. Vaughan \cite{vaughan}; Timmer \& Koenig \cite{timmer}; Benlloch et al. \cite{benlloch}).

\begin{figure}
\centering 
\resizebox{7cm}{!}{\includegraphics[angle=270]{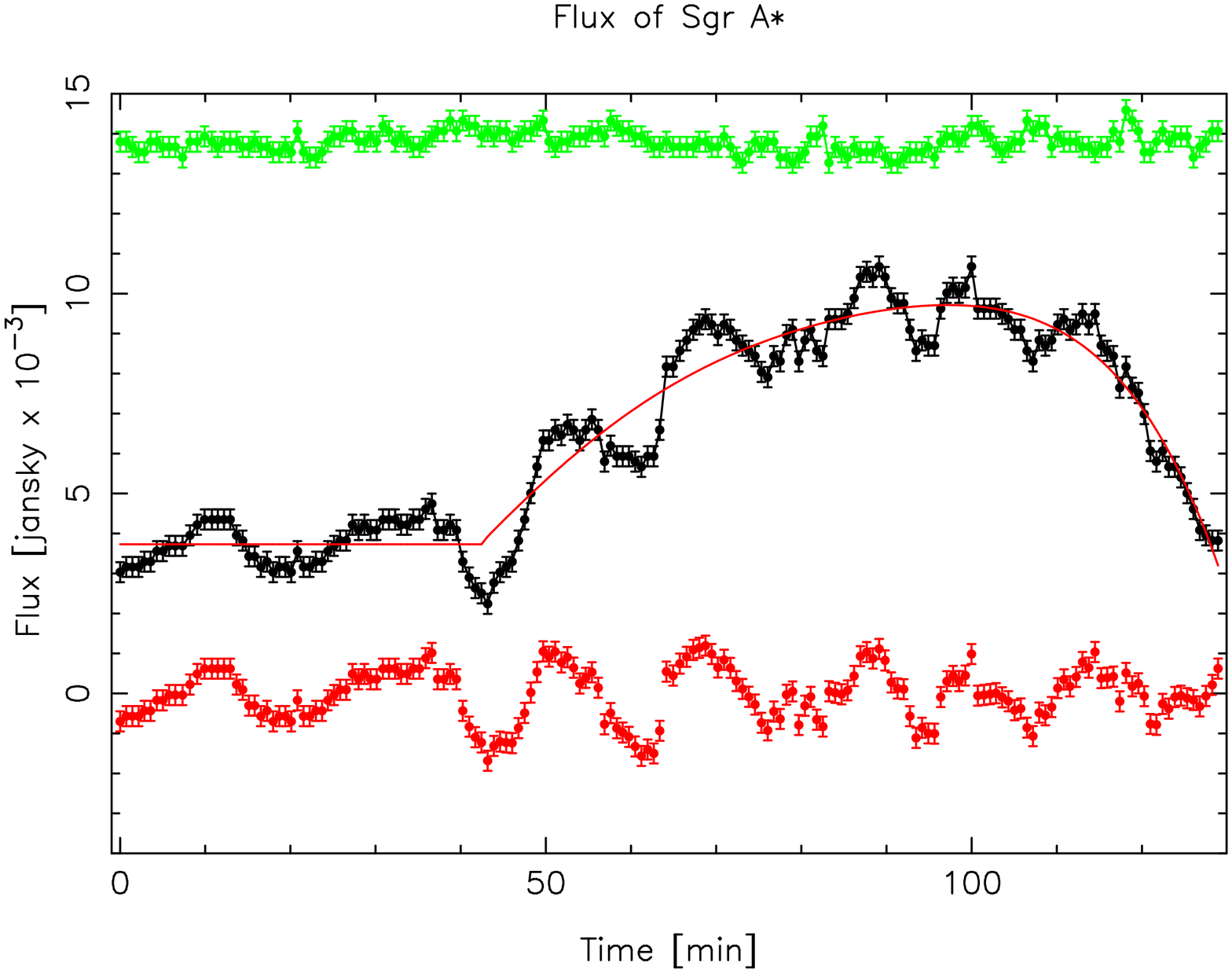}}

\resizebox{7cm}{!}{\includegraphics[angle=270]{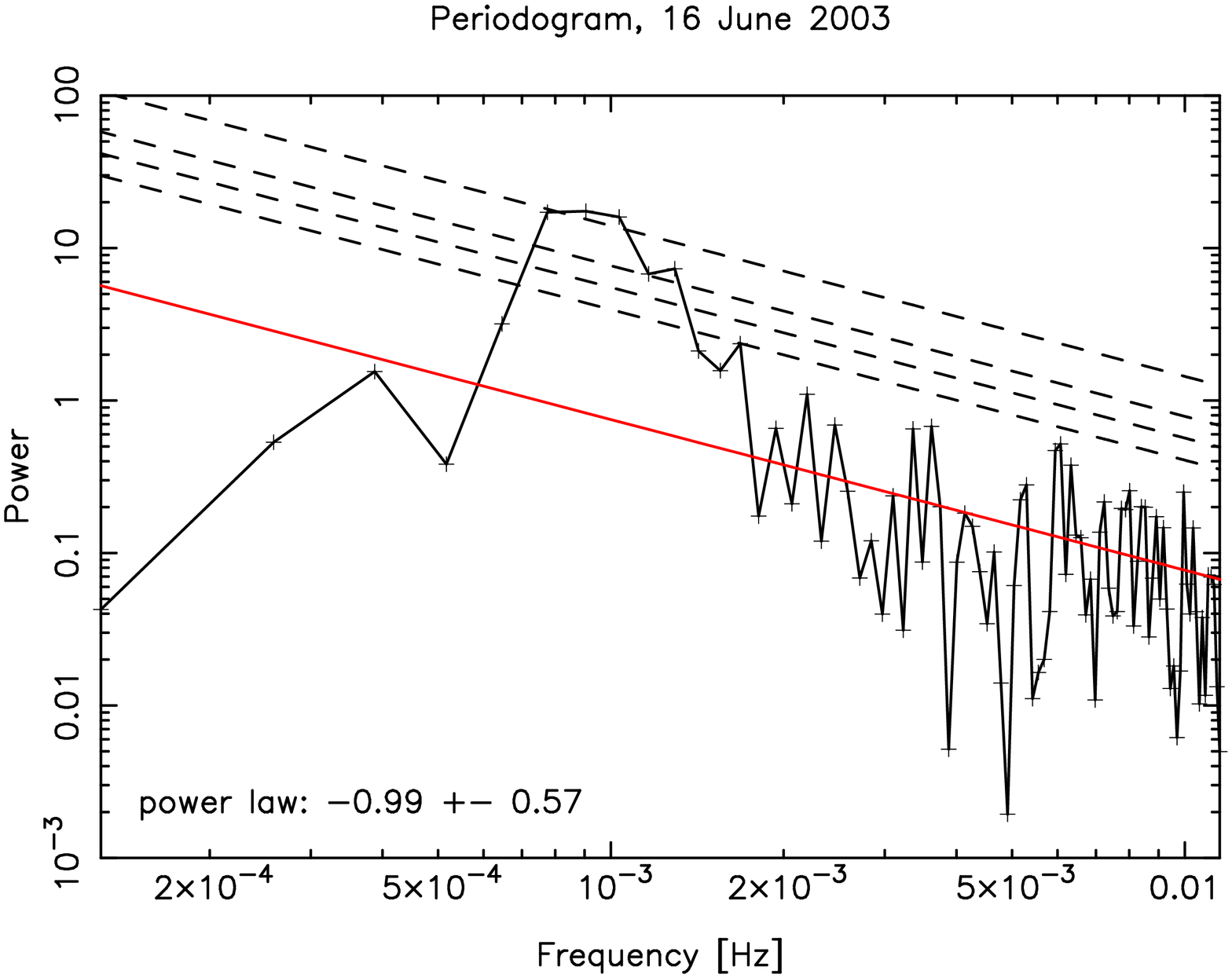}}

\caption{Sgr~A* NIR flare from 16 June 2003. Upper panel: Light curve
of Sgr~A* in black. A polynome fit to the overall flare and the residuals after subtracting this fit are shown in red. The light curve
of a constant star, S1, is shown in green.Lower panel: Periodogram of
the Sgr~A* flare after subtraction of the overall flare, i.e.\
corresponding to the lower, red curve in the upper panel. The straight
red line shows a de-biased power law fit to the data points when considering all
frequencies greater than $2\times 10^{-3}$\,Hz. {The dashed lines give the
1-, 2-, 3- and $5\sigma$ confidence bands derived with the method of Vaughan~(\cite{vaughan})}.  \label{Fig:flare2003}}

\end{figure}

In this section we re-analyse the so far best case for NIR quasi-periodicity
in Sgr~A*, that is the K-band flare from 16 June 2003 which was
originally reported by Genzel et al. (\cite{genzel}). Contrary to the other two mentioned flares, it has the advantage that apparent QPOs could be followed over 7 cycles.  The imaging data were
observed with the AO system/infrared camera NACO at the ESO VLT. The
detector integration time was 10\,s. Several images were averaged
before saving the data. The effective time resoltuion is about one
image every 40\,s. We reduced the data in a standard way (sky
subtraction, bad pixel correction, flat fielding). Applying a sliding
window, the median of 5 images at a time was constructed. From these
images, the PSF was extracted via point-source fitting with the
program \emph{StarFinder} (Diolaiti et al. \cite{diolaiti}). Subsequently, the
images were deconvolved with a Lucy-Richard deconvolution and beam
restored with a Gaussian beam. The flux of Sgr~A* and of stellar
sources in the field was extracted via aperture photometry with a
sufficiently large aperture to contain the entire flux of the
objects. Photometry calibration was done relative to several stars in
the field with a known flux. The resulting extinction corrected light curve (assuming $A_K=2.8$~mag) is shown in the
upper panel of Fig.~\ref{Fig:flare2003}. The black
curve shows the flux of Sgr~A* and the upper, green curve the flux of
S1, a star at $\sim0.4\arcsec$ distance from Sgr~A*. Its lightcurve is
assumed to be constant. The lower, red curve shows the light curve of
Sgr~A* after subtracting a constant (0-40 min) plus a 5th-order polynomial fit (40-130 min) to the overall flare.

The lower panel of Fig.~\ref{Fig:flare2003} shows a non-oversampled Lomb-Scargle periodogram of the
Sgr~A* flux after subtraction of the overall flare. There is a broad
peak visible around $2\times10^{-3}$\,Hz that corresponds to the
$16-21$\,min variability that can be seen in the light curve. {In order
to examine the significance of this peak, we followed the recipe of Vaughan~(\cite{vaughan})}. The assumed underlying power law is
indicated by the straight red line in the lower panel of
Fig.~\ref{Fig:flare2003}. This power law with an index of $-0.99$, i.e. a classical red noise, was
found by fitting all data points in the frequency range
$\geq 2\times10^{-3}$\,Hz. {The prescription of Vaughan~(\cite{vaughan}) takes the bias into account that originates when fitting the logarithm of the periodogram. Dashed lines in the plot of the periodogram
indicate the 1-, 2-, 3- and $5\sigma$ confidence bands which can be calculated analytically. We also used Monte Carlo simulations to derive these confidence limits which leads to the same result.

It can be seen that the peak around $10^{-3}$\,Hz reaches the $5\sigma$ threshold. Even after correcting for the model uncertainty, as it is suggested in Vaughan~(\cite{vaughan}), the significance is still at the $4.2\sigma$ level.  

The critical point in this procedure is the choice of the slope of the
power spectrum. We followed the prescription of Vaughan~(\cite{vaughan}). We think our approach is justified because the periodogram shows a clear component of red noise (see also discussion in Belanger et al.~\cite{belanger}).} 

We conclude
that, on the one hand, there are strong indications for a
quasi-periodicity in the NIR flare from 16 June 2003. On the other
hand, this single flare by itself is not enough to present ironclad
evidence. However, two more events were reported that support the presence of a
periodicity in the considered time range: the second K-band flare
reported by Genzel et al. (\cite{genzel}) and the flare observed in
polarimetry by Eckart et al. (\cite{ecki2}). In fact, everytime sub-flares have been observed so far, they showed a period that lies within the $20\pm 3$\,min interval. The likelihood of seeing sub-flares at this interval in consecutive observations is small. The probabilities multiply and can be estimated simply: the probability to detect a peak in the periodogram of a single observation is at most at the percent level, see discussion above. That means that the chance to observe a peak at the same frequency interval in three consecutive runs is $\sim 0.1\%$, and $\sim 0.01\%$ if a fourth NIR flare showing sub-flares has a period at $\sim 20$\,min. Therefore we believe that the evidence for an intrinsic cause of these modulations must be taken serious.

\section{The Hot Spot/Ring Model} \label{spot}

Disk accretion is generally thought to be the source of galactic nuclear
luminosity and the driver of activity whenever the angular momentum
dominated disk-type accretion takes place. In the case of Sgr~A*, the
accretion rate onto the central black hole and the net angular momentum
of the accreted gas appear to be very small
($\simeq10^{-7}M_{\sun} \;\mbox{yr}^{-1}$, see Bower et al. \cite{bower}) and a persistent accretion disk
is not present {(Coker \& Melia~\cite{coker})}. Despite this fact, the very occurrence of the flares and
the minute time-scale of their rise and decay indicate that the
immediate neighbourhood of the black hole is not entirely empty; likely,
episodic accretion takes place. In fact, Moscibrodzka et al. \cite{moscibrodzka}
suggest that a transient torus develops at a distance of a few or a few
ten gravitational radii and exists for several dynamical
periods. The inner disk/torus is subject to violent instabilities, but
it is difficult to identify the dominant mechanism responsible causing
its destruction.  This makes it
tempting to attribute the outbursts to some type of disk
instability, for instance of the limit-cycle type as known in dwarf
novae (albeit with the possibility of a physical cause different
from the hydrogen ionization/recombination there). Disks in the
vicinity of black holes live in a relativistic environment, and
proper modeling requires relativistic \mbox{(magneto-)hydro}dynamics ({see e.g. Tagger \& Melia \cite{tagger})}. In
the following, however, disk instabilities can be ruled out by such
a wide margin that a non-relativistic order-of-magnitude treatment
suffices our purpose.

So close to the black hole, self-gravity will not play a role,
allowing for the standard one-zone approximation of the vertical
hydrostatic equilibrium, which relates the azimuthal velocity
$v_\varphi$, the sound velocity $c_\mathrm{s}$, the
(half-)thickness of the disk $h$, and the radius $s$ by

\begin{equation}
\frac{h}{s} = \frac{c_\mathrm{s}}{v_\varphi}.
\end{equation}

The rise time $\Delta t_+$ of a disk outburst may be estimated by
the radial range of the disk involved in the outburst, $\Delta s$,
and the propagation speed of the outburst, which is $\alpha
c_\mathrm{s}$ (Meyer \cite{meyer}; {see also Liu \& Melia~\cite{liu}}),

\begin{equation}
\Delta t_+ \approx \frac{\Delta s}{\alpha\ c_\mathrm{s}} \ge
\frac{\Delta s}{c_\mathrm{s}} = \frac{\Delta s}{v_\varphi}
\frac{s}{h} \approx t_\mathrm{dyn} \frac{s}{h} \frac{\Delta s}{s}
\end{equation}
with the dynamical timescale $t_\mathrm{dyn} \approx
\frac{s}{v_\varphi}$. Here $\alpha$ is a parameter
related to the disk viscosity and is of order unity or smaller. For the outburst to be a disk phenomenon and
not just something happening in a small ring, it is required $\Delta
s \gg s$.

Even for an only moderately thin disk of, let's say, $h/s =
10^{-0.5}$ this then requires rise times which are more than an
order of magnitude longer than the dynamical time scale. The
smallest dynamical time scale allowed for is the one at the ISCO. As discussed above, however, the entire
outburst (rise plus decline) encompasses only a few dynamical time
scales.

This ratio of the observed time scales makes a {\it normal\/} disk
instability an exceedingly unlikely cause of the outbursts. It
furthermore shows that, at best, a relatively small radial range of
the disk ($\Delta s \sim s$), close to the ISCO,
may be involved in whatever causes the outbursts.

It is, however, noteworthy that, for a moderately thin accretion
disk, the duration of the outburst is of the order of the timescale
a sound wave needs to travel around the black hole once. This makes
a scenario attractive in which the outburst is caused by some
non-axisymmetric process (be it a disk
instability of a non-axisymmetric type, confined to the innermost disk
regions, be it material falling into the disk there locally), and a
sound wave originating from this event and traveling the inner disk
until the outburst comes to an end when the actual source of it has
been accreted into the black hole. While the sub-flares tracing the
actual local event disappear suddenly as soon as that region has
been accreted, the outburst itself ceases somewhat slower as the
soundwave, of course, travels not only azimuthally, but also
radially outwards, though only a small distance as shown above.
%

Motivated by these considerations, we adopt a two component hot spot/ring model in order to fit the data presented in Eckart et al. (\cite{ecki2}). In this model the broad underlying flare is caused by a truncated disk/ring. The sub-flares are due to a compact emission region on a relativistic orbit around the MBH. Such hot spots can be created in reconnection events, in which thermal electrons are accelerated into a broken power law (Yuan et al. \cite{yuan2}). General relativistic effects imprint on the synchrotron radiation of such inhomogeneities (Broderick \& Loeb \cite{broderick1}, \cite{broderick2}; Dovciak et al. \cite{dovciak}; {Hollywood et al.~\cite{hollywood}; Hollywood \& Melia~\cite{hollywood2}; Bromley et al.~\cite{bromley}; Falcke et al.~\cite{falcke}}; Pineault \cite{pineault}). Redshifts, lensing, time delay, change in emission angle and change in polarization angle belong to these. The code by Dovciak et al. (\cite{dovciak}) takes these special and general relativistic effects into account by using the concept of a transfer function (Cunningham \cite{cunningham}). A transfer function relates the flux as seen by a local observer comoving with an accretion disk to the flux as seen by an observer at infinity. This transfer of photons is numerically computed by integration of the geodesic equation. For the change of polarization angle the method of Connors \& Stark (\cite{connors}) has been used. In geometrical optics approximation, photons follow null geodesics and
their propagation is not affected by the spin-spin interaction with a
rotating black hole (Mashoon \cite{mashhoon}). This means that wave fronts do not
depend on the photon polarization, and so the ray-tracing through the
curved spacetime of the black hole is adequate to determine the observed
signal.

{Following the above discussion we have fixed the radial extent of the disk to $2R_S$, beginning at an inner edge. We realize that there is a range of definitions for an inner edge
that depend on the accretion rate and may be non-axisymmetric and
time-variable (Krolik \& Hawley \cite{krolik}).
They are, however, generally located in the vicinity of the
marginally stable orbit. For simplicity we have assumed that for $a_\star \ga 0.5$ both coinside,
and our general result -- that the observed time variability is compatible with
a discription of a spot/ring combination orbiting a central MBH --
does not depend strongly on this assumption. Due to the observed timescale of the QPOs, the spot is within the marginally stable orbit for $a_\star \la 0.5$. In this case we assumed the spot to be freely falling and the disk to have its inner edge at $\sim R_S/2$ above the event horizon. Magnetic field effects could bar the spot from freely falling (Krolik \& Hawley~\cite{krolik}) but require full relativistic MHD simulations that are beyond the scope of this work.  }

The disk/ring's time behaviour was assumed to be Gaussian and to account for the main flare. The spot is orbiting within the disk and the equatorial plane of a Kerr BH. Its intrinsic luminosity follows a two-dimensional normal distribution with $\sigma \sim 1.5 R_S$, but being cut off in the radial direction to fit within the disk. The spot follows a circular trajectory ($a_\star \ga 0.5$) chosen such that the observed periodicity is matched. Note that the radius for the corresponding orbit is spin dependent, see equation (\ref{period}) or Bardeen et al. (\cite{bardeen}). {In the case $a_\star \la 0.5$ the spot is freely falling and it has to be checked whether the timescale can be matched within its uncertainty.}

We identify all three sub-flares with the same spot. This can well be matched with the corresponding synchrotron cooling timescale (Eckart et al. \cite{ecki1},b; Yuan et al. \cite{yuan1}, \cite{yuan2}). The observed differences in the flux of each sub-flare are very likely due to intrinsic changes of the spot, i.e. the spot is very likely to evolve. Regarding the fact that the physics of such an inhomogeneity in the accretion flow is unknown at present, we simply model the changes of the intrinsic luminosity by hand. This is done by changing the value for the spot luminosity discontinuously instead of some unphysical parametrization. {Normalized to the third revolution ($\sim 36-54$\,min) the spot has an intrinsic luminosity of 50\% between 0-18\,min, 20\% between 18-36\,min and 10\% after 54\,min. } The small jumps seen in some light curves are an artifact of this procedure.

Another approach would be to assign different spots to different sub-flares. Assuming a typical spot-lifetime of approximately one revolution and that every spot is on roughly the same radius, multiple single spots can  reproduce a peak in the periodogram (Schnittman \cite{schnittman}). This procedure could naturally lead to better fits, because the phase of each spot would be a free parameter for every single sub-flare. We rejected that approach, however, as it only increases the already large number of degrees of freedom.

To investigate the parameters of Sgr A* we have fitted for the inclination angle, the dimensionless spin parameter $a_\star$, the brightness excess of the spot with respect to the disk, the initial phase of the spot on the orbit, the orientation of the equatorial plane on the sky and the polarization degree of the disk (constrained to $\leq 15\%$) and the spot ($\leq 70\%$, a value that can well be achieved via synchrotron radiation).  As the last five parameters can be changed after the ray tracing, it costs little computational effort to find the least $\chi ^2$ values for given spin and inclination (Lourakis~\cite{lourakis}). {For these two parameters we have chosen a discrete grid with $0 \leq a_\star \leq 1$ and steps of 0.1 and an inclination angle $10^\circ \leq i \leq 70^\circ$ with steps of $5\degr$; $i=0\degr$: face-on.} For $i > 70\degr$ multiple images could become important, which are not included in our treatment (see next Sect.). 

As the observational data include the effects of foreground polarization, we also take these into account. Following the calibration of Eckart et al. (\cite{ecki2}) we fix it to be 3.4\% at $25^\circ$.  As only rough estimates can be given on the magnetic field of the spot, we did the fits for two different field configurations as an approximation  (Pineault \cite{pineault}). The first configuration is such that the resulting projected E-Vector is always perpendicular to the equatorial plane. Such preferred orientation could result from perturbations in the disk similar to sunspots (see also Shakura \& Sunyaev~\cite{shakura}).   As a second configuration we have allowed for a global azimuthal magnetic field. This behaviour may be caused by the magneto-rotational instability and is motivated by global MHD simulations (Hirose et al. \cite{hirose}). Contrary to the first case, this configuration leads to a rotation of the $E$-vector along the orbit from a Newtonian perspective (see Figure~2).

\begin{figure}[t]
\resizebox{\hsize}{!}{
\includegraphics{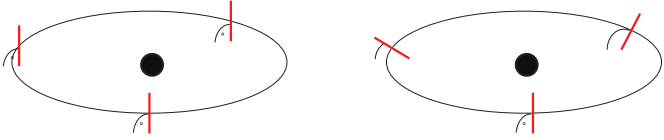}}
\label{car}
\caption{Sketch of the Newtonian behaviour of the $E$-vector (red) for the two magnetic field configurations. The black circle represents the MBH, the ellipse within the equatorial plane indicates a projected Keplerian orbit of a hot spot in the local comoving frame. Left: the case where the $E$-vector is always constant. Right: the case of an azimuthal magnetic field (like the ellipse) where the $E$-vector of the emitted synchrotron radiation rotates.}
\end{figure}

\section{Results \& Discussion} \label{concl}

\begin{figure}[t]
\centering
\resizebox{7cm}{!}{\includegraphics[angle=270]{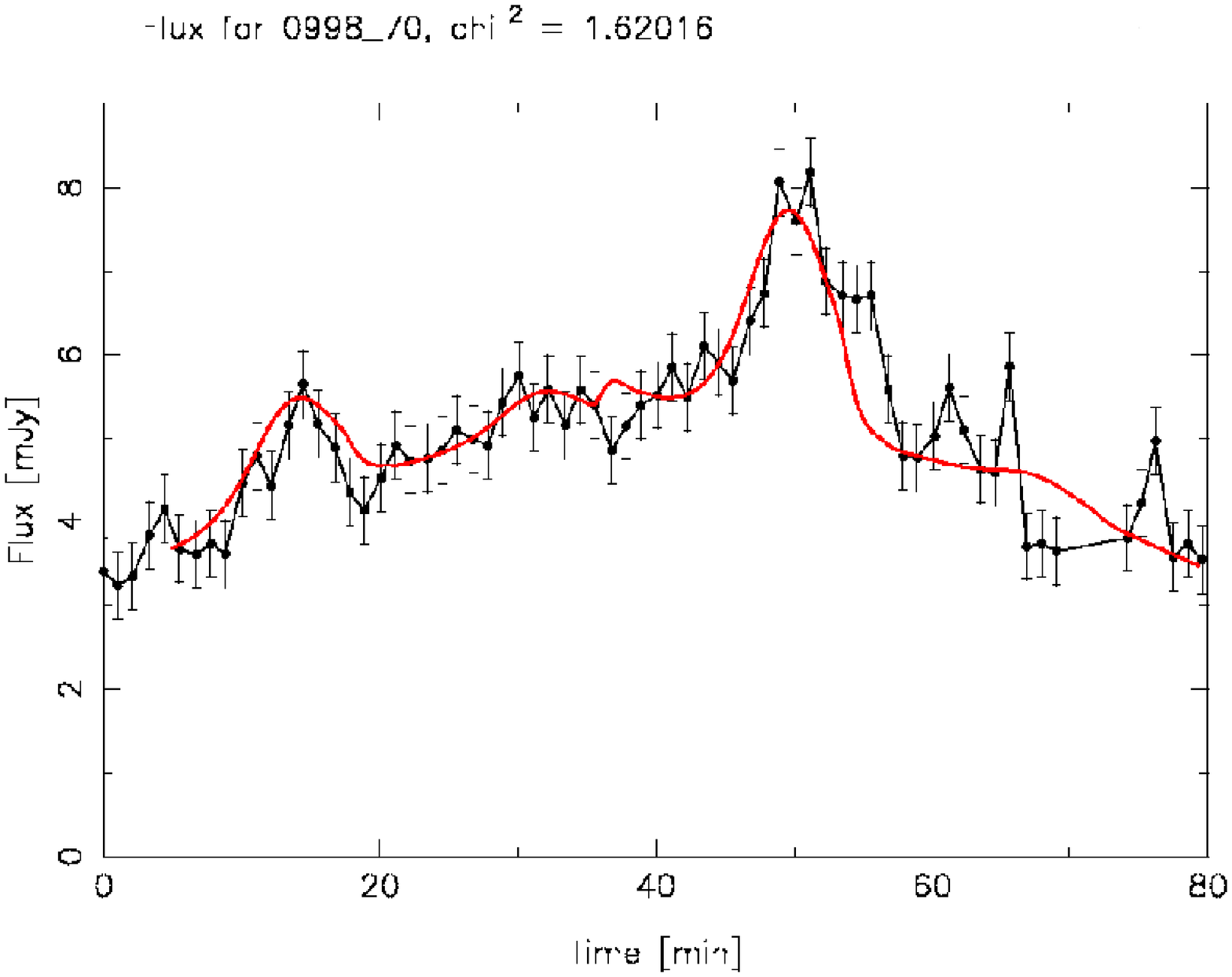}}

\resizebox{7cm}{!}{\includegraphics[angle=270]{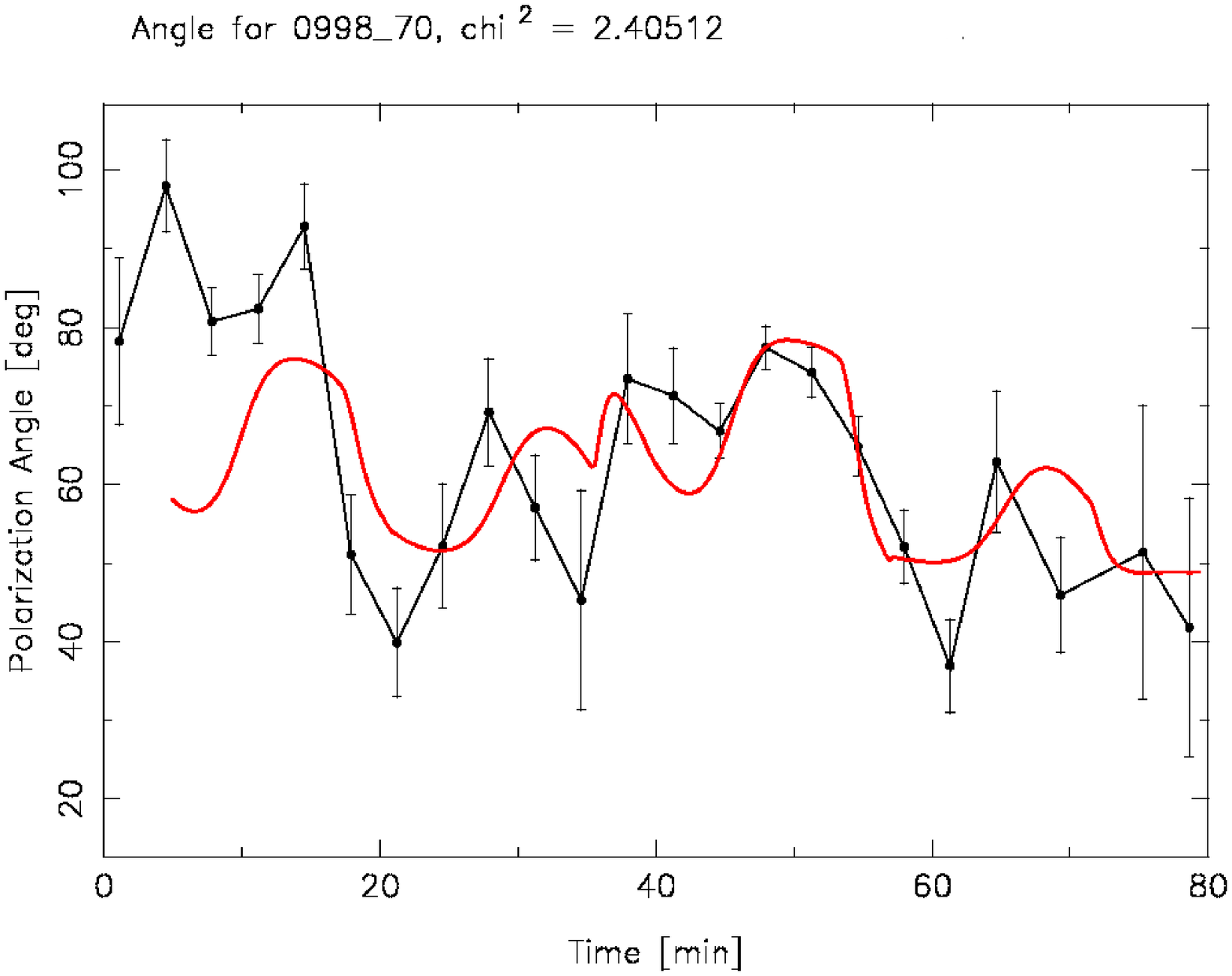}} 

\resizebox{7cm}{!}{\includegraphics[angle=270]{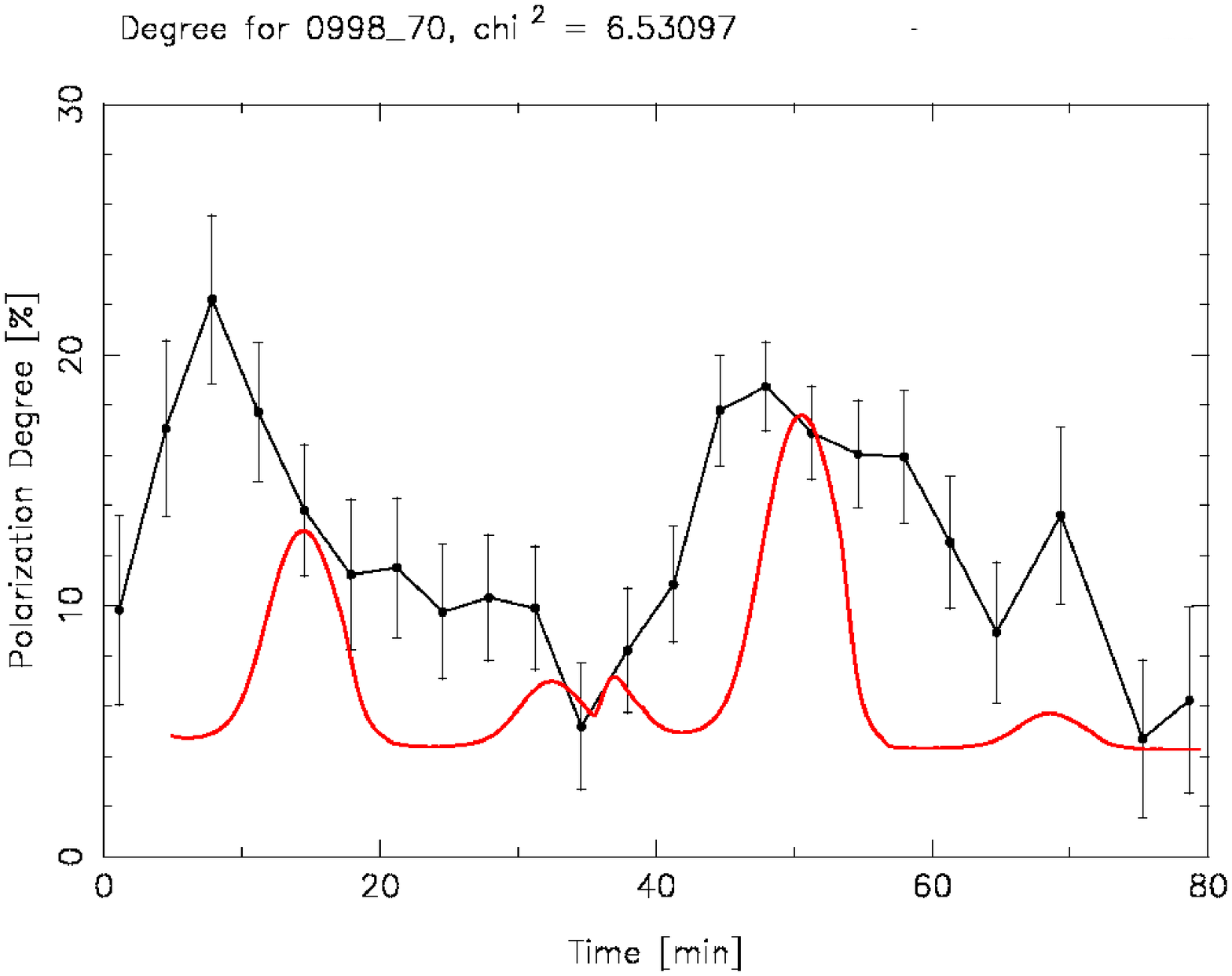}} 
\caption{The best fit solution (in red) for a constant $E$-vector. Shown is the flux (top), polarization angle (middle), and polarization degree (bottom). The parameters of the model are $i = 70\degr$, $a_\star \approx 1$, a location of the projection of the disk on the sky of $100\degr$ and a polarization degree of the disk (spot) of $4\%$ ($60\%$). The spot is orbiting at a \mbox{radius $r=4.4 r_g$.}}
\label{fig1}
\end{figure}

\begin{figure}[t]
\centering
\resizebox{7cm}{!}{\includegraphics[angle=270]{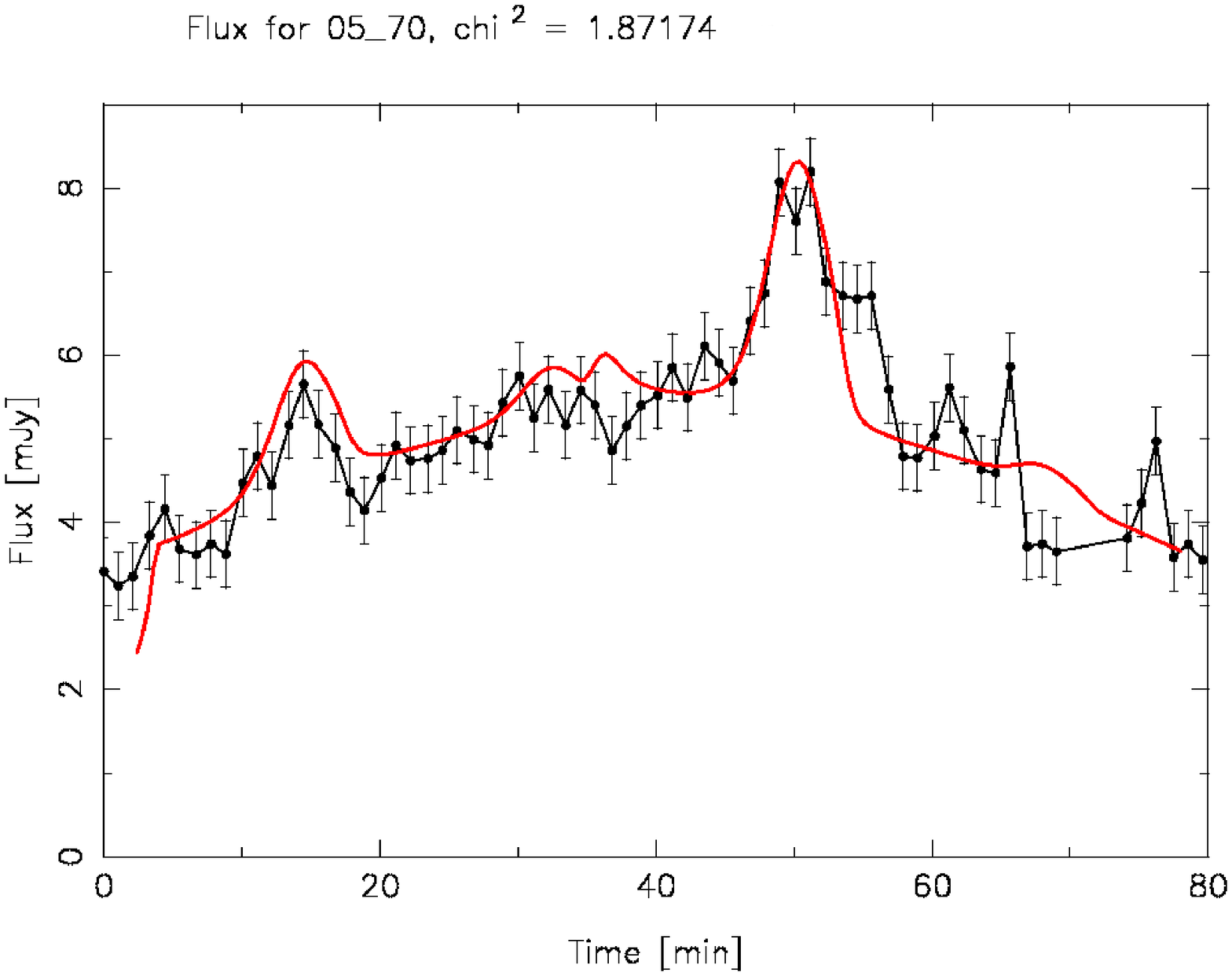}}

\resizebox{7cm}{!}{\includegraphics[angle=270]{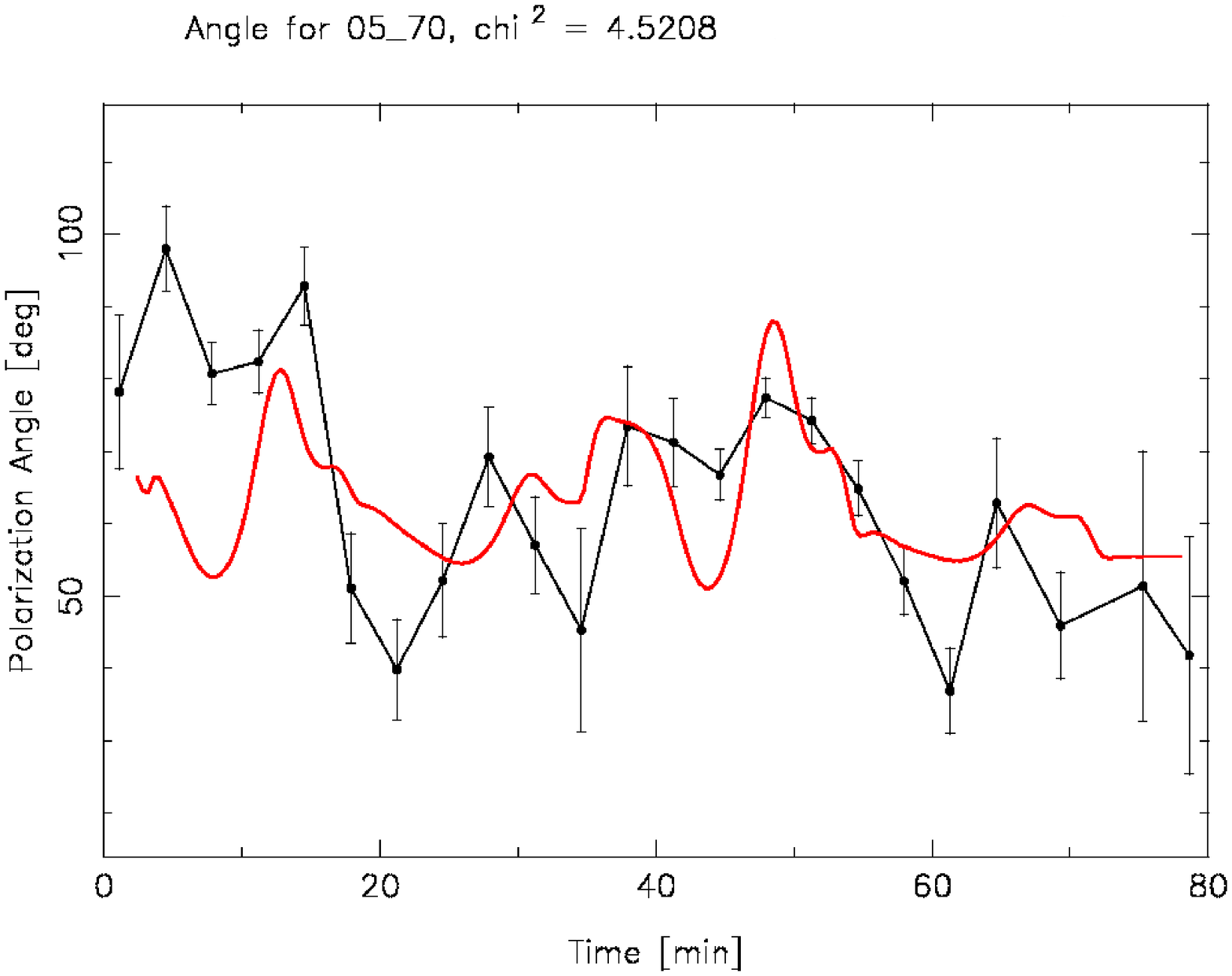}} 

\resizebox{7cm}{!}{\includegraphics[angle=270]{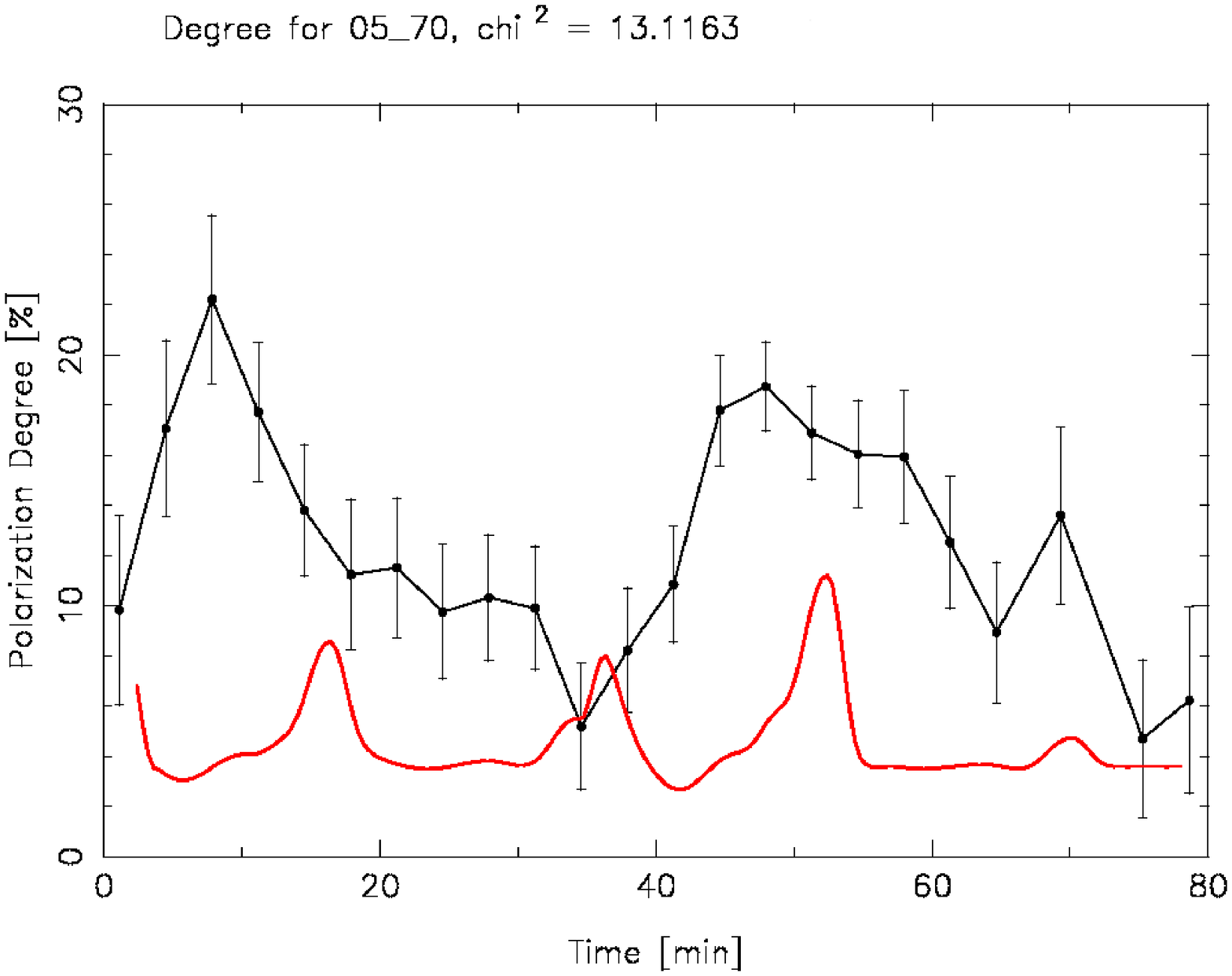}}
\caption{The best fit solution (in red) for a global azimuthal magnetic field. Shown is the flux (top), polarization angle (middle), and polarization degree (bottom). The parameters of the model are $\textnormal{inclination} = 70\degr$, $a_\star \approx 0.5$, a location of the projection of the disk on the sky of $0\degr$ and a polarization degree of the disk (spot) of $9\%$ ($60\%$). The spot is orbiting at a \mbox{radius $r=4.5 r_g$.}}
\label{fig2}
\end{figure}

Figure \ref{fig1} shows the fit with the least $\chi^2$ value within the discrete grid discussed in Sect.~\ref{spot} in case of a constant $E$-vector. High inclination and high spin give the best solutions. The case of the azimuthal magnetic field can be seen in Figure \ref{fig2}. Here high inclination and medium spin is preferred. Within the hot spot model the degree of polarization does not necessarily have to be fitted. The spot may have influence on the disk and may polarize the disk flux.

\begin{figure}[t]
\centering
\resizebox{7cm}{!}{\includegraphics[angle=270]{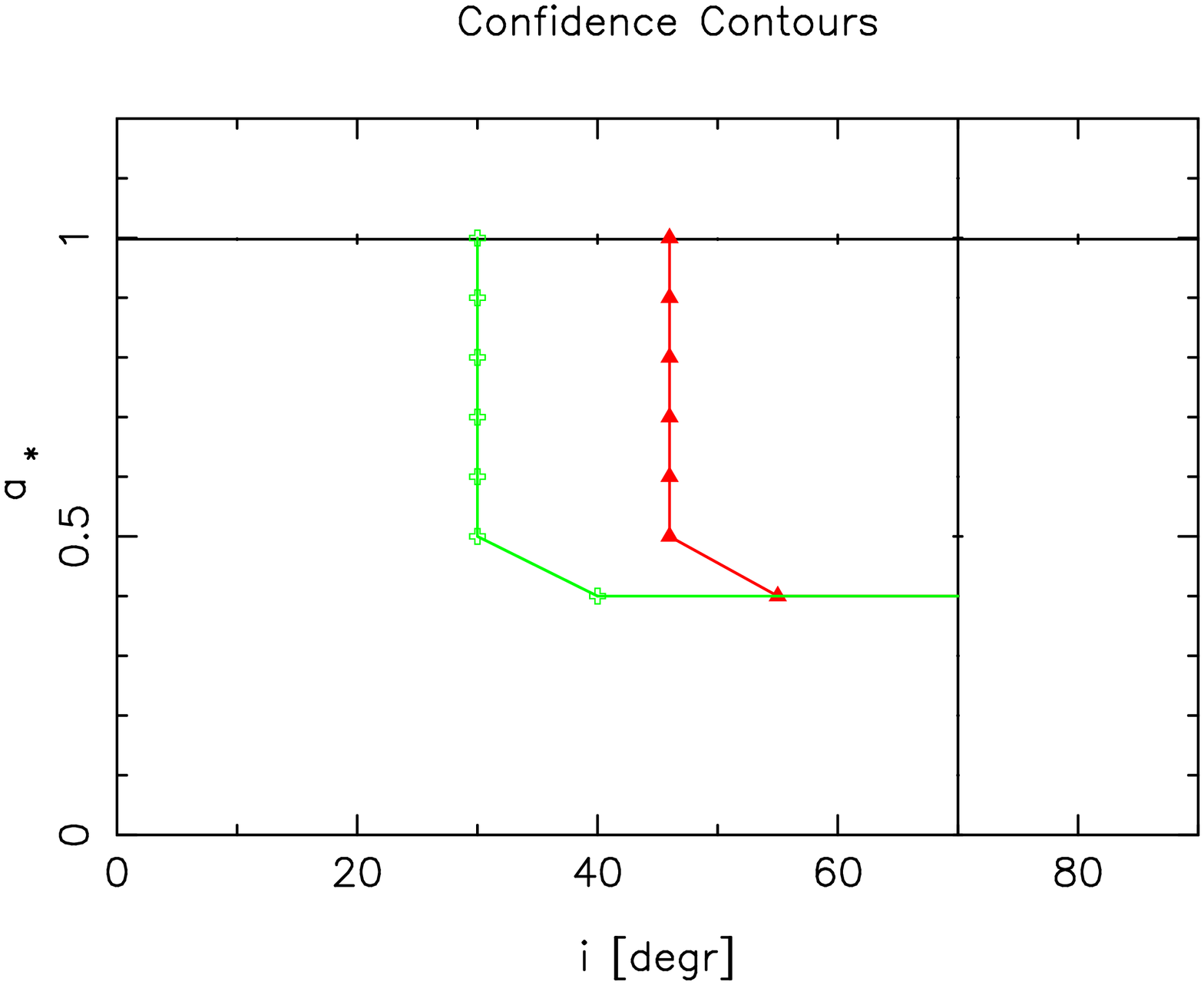}}

\resizebox{7cm}{!}{\includegraphics[angle=270]{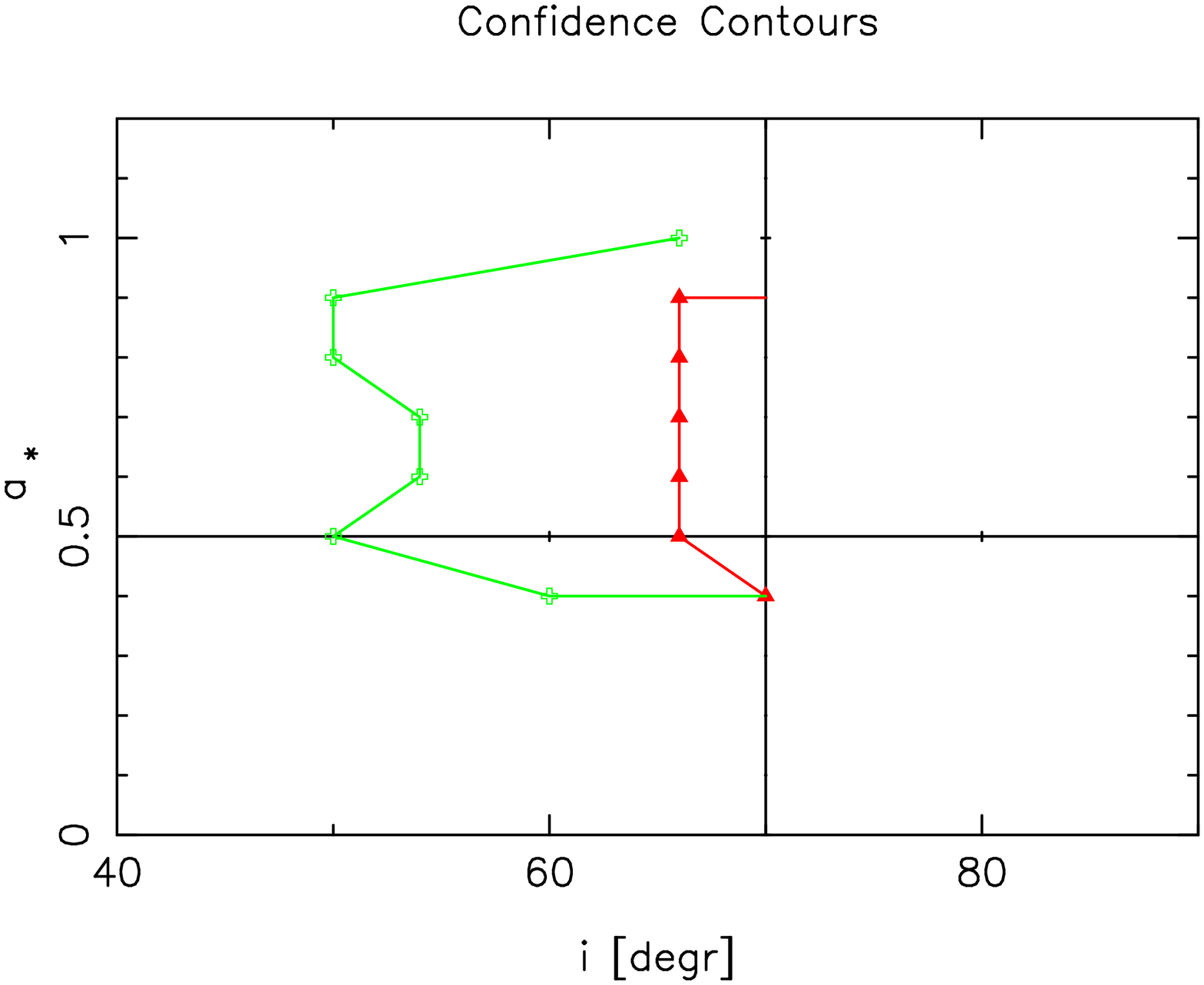}} 
\caption{The confidence contours for the constant $E$-vector (top) and the azimuthal magnetic field case (bottom). The red (green) line is chosen such that the projection onto one of the parameter axes gives the $1\sigma$ ($3\sigma$) limit for this parameter. The respective $\chi^2$-minimum is marked by the big cross. Note the different scales on the abscissae.}
\label{confcon}
\end{figure}

{The confidence contours shown in Figure~\ref{confcon} reveal that for both magnetic field cases the spin paramter $a_\star$ cannot be well constrained from the current data with its uncertainty. Within the $3\sigma$ limit $a_\star$ can be inferred to lie in the range $0.4 \leq a_\star \leq 1$. The range of the inclination is different for the two magnetic field scenarii. As the $\chi^2$-minimum is lower for the constant $E$-vector, this case should be weighted more. That means the inclination is $\ga 35\degr$ on a $3\sigma$ level.}

Our solution with $a_\star \ge 0.4$ is in agreement with Genzel et al. (\cite{genzel}) and Belanger et al. (\cite{belanger}), who inferred values of $a_\star \geq 0.5$ and $a_\star \geq 0.22$, respectively. They arrived at these lower limits by interpreting the peak in the power spectra of their measured light curves using the hot spot model. As it is not possible to decide where the spot is orbiting, only a lower bound can be given with the equality applying if the spot is exactly on the ISCO. Only polarimetric observations together with a relativistic modeling can improve on this within the spot model due to the additional indepent information. {Our analysis of recent polarimetric measurements, however, shows that very accurate data are needed to give strong constraints on the spin parameter.} 

One should be careful about the role of general relativistic effects for
polarization fluctuations and there does not seem to be a
clear way towards a unique interpretation at the present stage. Likewise the inferred values of $a_\star$ and the inclination are
subject to many uncertainties -- mainly concerning the
geometry of the source and its intrinsic polarization.

We note that we did not take higher order images into account. While they offer the unique possibility of testing general relativity in a strong field quantitatively (Broderick \& Loeb \cite{broderick2}), their detection is extremely difficult and lies well beyond present observational capabilities. Interferometry in the NIR or sub-mm domain may resolve this question in the future. Only for special geometries, where strong lensing is very important, they yield a non-negligible contribution (Broderick \& Loeb \cite{broderick2}; Bozza \& Mancini \cite{bozza}; Horak \& Karas \cite{horak1}, \cite{horak2}). The lack of multiple images in our models limits their applicability to $i \la 70\degr$. One may, however, speculate, what signatures of multiple images to look for. Time delay between 1st and 2nd images is a very characteristic number, related to BH mass: it is given by the circumference of the photon circular orbit,
$R_p=2(1+\alpha)GM/c^2$, where
$\alpha=\cos[\frac{2}{3}\arccos\,a_{\star}]$ (Bardeen et al. 1972). Hence, the expected delay is of
the order of $575\left( M/(3.6\times10^6M_{\sun}\right)$~sec in the Schwarzschild
case ($a_{\star}=0$; this interval becomes roughly a factor of 3 shorter
in case of a maximally rotating black hole and prograde orientation of
the photon trajectory, while it is about 1.3 times longer for the
retrograde orientation). 
 Unfortunately, the second image provides an order of magnitude less
photons, and higher order (indirect) images are still weaker because the
observed radiation flux decreases exponentially with the image order $n$
increasing (its observed flux is roughly proportional to
$\exp(-2\pi{n})$; cf. Luminet \cite{luminet}), unless, again, a special geometry
is assumed in which caustic lensing occurs.

In fact, the importance of significant lensing enhancement of the
observed flux from accretion disks has been traditionally neglected,
because it would require large inclination angles. Such geometry is not
likely in case of standard (optically thick) accretion disks, because
self-obscuration does not allow to observe the indirect images. However,
the situation is different for optically thin flows (as noticed by Bursa
et al. \cite{bursa}). Therefore, Sgr A* is a suitable object in which the flux
fluctuations could be considerably modulated by strong-gravity lensing if the inclination of the putative disk is very large {(see also Falcke et al.~\cite{falcke}; Bromley et al.~\cite{bromley})}.

In case of non-zero rotation of the BH, the expected lightcurves exhibit stronger
fluctuations because the structure of light-ray caustics is more complex
(e.g. {Hollywood \& Melia~\cite{hollywood2}}; Rauch \& Blandford \cite{rauch}; Viergutz \cite{viergutz}). Rapid decay of the signal
strength with increasing $n$ holds also for a rotating black hole,
although the prospect of detecting the higher-order images appears even
more tempting, because their mutual delays at the point of arrival and
relative polarization degree can set tight constraints on the black-hole
angular momentum. In conclusion, higher order images are not possible to measure yet, though it would be very useful and should be attempted in the future because their interpretation could be less ambiguous.

\section{Summary}

After showing that the quasi-periodicity of the June 2003 flare (Genzel et al. \cite{genzel}) appears to be clearly significant, we have outlined a consistent physical picture for the NIR flares of Sgr~A*. In that model the broad flares with a typical timescale of 60 - 100 min are due to a sound wave that travels around the MBH once, caused by a non-axisymmetric perturbation in the innermost region of an accretion disk. The sub-flares superimposed on the broad flare are due to an orbiting compact source of transiently heated and accelerated electrons (Yuan et al. \cite{yuan2}), whereas the acceleration mechanism may be linked to the local event causing the sound wave. As we have shown, modeling this with the simple hot spot idea leads to reasonable fits of the polarimetric NIR observations presented by Eckart et al. (\cite{ecki2}). The least $\chi^2$ values are obtained for a spin parameter $a_\star$ close to unity, {on a $3\sigma$ level it is constrained to the range $0.4 \le a_\star \le 1$. The inclination tends to be high, it can be constrained to $i \ge 35\degr$.}

\begin{acknowledgements}
We want to thank Avery Broderick for enlightening discussions during the GC 2006 meeting. The anonymous referee made comments that helped to improve the paper substantially. We are grateful to Thomas M\"uller, Frank Grave, Andrei Lobanov and Michael Nowak for their support and advice and especially to Claus Kiefer for the promotion of this project. A.E. and R.S. thank Reinhard Genzel and the whole MPE infrared goup for fruitful discussions. L.M. is supported by the International Max Planck Research School (IMPRS) for Radio and Infrared Astronomy at the Universities of Bonn and Cologne. 
\end{acknowledgements}

\end{document}